\documentclass{article}

\usepackage[english]{babel}
\usepackage{natbib}
\usepackage{bm}
\usepackage{listings}
\usepackage[export]{adjustbox}
\usepackage{subfloat}

\usepackage[letterpaper,top=2cm,bottom=2cm,left=3cm,right=3cm,marginparwidth=1.75cm]{geometry}

\usepackage{amsmath}
\usepackage{graphicx}
\usepackage[colorlinks=true, allcolors=blue]{hyperref}

\title{Approximate Shapley value estimation using \\
sampling \textit{without} replacement\\
and variance estimation via the new \\
Symmetric bootstrap and the Doubled half bootstrap }

\author{Fredrik Lohne Aanes}

\begin{document}
\maketitle

\begin{abstract}
In this paper I consider improving the KernelSHAP algorithm. I suggest to use the Wallenius' noncentral hypergeometric distribution for sampling the number of coalitions and perform sampling without replacement, so that the KernelSHAP estimation framework is improved further. I also introduce the Symmetric bootstrap to calculate the standard deviations and also use the Doubled half bootstrap method to compare the performance. The new bootstrap algorithm performs better or equally well in the two simulation studies performed in this paper. The new KernelSHAP algorithm performs similarly as the improved KernelSHAP method in the state-of-the-art R-package shapr, which samples coalitions with replacement in one of the options. \end{abstract}

\section{Introduction}

In the classical KernelSHAP \citep{Lundberg2017} framework the goal is to lower the computational cost of Shapley value estimation. This is done by not calculating the contribution fucntion for all coalitions. Instead one samples a subset of coalitions typically \textit{with} replacement and use the Shapley weights in the procedure, so that coalitions with larger weight have a higher probability of being included in the sample compared to coalitions with a lower weight. The sampling frequencies are used in the calculations. In \cite{olsen2025improving} they improve the method by using expected weights instead of the sampling frequencies (as in the original procedure). Sampling in this new procedure is still performed \textit{with} replacement. The method is implemented in the R-package shapr \citep{jullum2025shapr} on CRAN. I use version 1.0.7 of this package. 

In my new work I use existing methods and replace some components in the original KernelSHAP procedure so that sampling can be done efficiently without replacement. Experiments show that the new method has similar performance to the new method in shapr. 

One major benefit of the new method is that finite population methods can be used to obtain variance estimates of the approximated Shapley values. I introduce a new, easy way of obtaining bootstrap estimates when sampling is simple random sampling without replacement. I note that when all possible coalitions are accounted for when computing SHAP-values, there is no uncertainty in these values. The traditional bootstrap algorithm can be applied with ease when all or almost every coalition has been accounted for. However, the true variance of each Shapley value should be relatively small, due to the finite population constraint. This constraint is not necessarily fulfilled when applying ordinary bootstrapping, since the bootstrap ``iid'' sample assumption is not fulfilled. As previously explained, when we have sampled and calculated the contribution function for each coalition, we have the correct Shapley values with 0 in variance. Some would argue that different estimation techniques give different estimates of the contribution function. This is correct, and we fix the estimation technique of the contribution function when we use the KernelSHAP framework. This observation holds for all the estimation methods that use sampling of coalitions, and not only my method. 

The paper is structured as follows. In Section \ref{sec:FinitePop} I briefly present finite population theory, also known as Survey theory. In Section \ref{sec:BriefOverview} I provide a short overview of Shapley values for explaining model predictions. In Section \ref{sec:CombineFiniteAndKernelSHAP} I suggest how finite population theory can be applied within the KernelSHAP framework, so that sampling without replacement can be performed. In Section \ref{sec:VarEstimation} I consider variance estimation of Shapley values. I consider two valid estimators for estimating the standard deviation of SHAP-values obtained through the new computational algorithm. One of them is well-known, the Doubled half bootstrap of \cite{antal2014new}, while the second one, the Symmetric bootstrap, is a new method I develop in this paper. I also propose an invalid, but intuitive method of obtaining the standard deviations. I will explain why it is invalid and therefore I will not use it in the simulation studies. In Section \ref{sec:SimStudies} I perform two simulation studies on the Life Expectancy (WHO) Fixed dataset from Kaggle, one with a smaller subset and one with a larger subset of covariates. I also discuss the results in this section. Finally in Section \ref{sec:conclusion} I provide a brief general discussion and a conclusion.   

\section{Finite population/Survey theory}\label{sec:FinitePop} In finite population theory, we have a universe with finite number of elements. I provide a short introduction where I consider simple random sampling without replacement. For a more detailed and general introduction see e.g the influential book \cite{cochran1977sampling} or, for more recent work, see \cite{tille2020sampling}.

I consider a population with units/members indexed with 1,2,3 and so forth up to $N$. I then sample the population and obtain a sample $S$ of size $n_S$. I sample without replacement. The first order inclusion probabilities only consider the univariate inclusion probabilities, where inclusion means the probability that a unit is in the final subsample. The order in which units are sampled is not important. Only membership/non-membership is important. Let $\pi_i$ be the first order inclusion probability that unit $i$ is in sample $S$. The second order inclusion probability $\pi_{i,j}$ is the probability that both $i$ and $j$ is in $S$. In the current work I consider simple random sampling without replacement, for which it is known that $\pi_i=n_S/N$ and $\pi_{i,j}=n_S (n_S-1)/(N (N-1))$ and $i\neq j$. I note that $\pi_{i,i}=\pi_i$. 

Finite population inference theory, also known as design based inference theory, should be compared with model based inference theory, in which a superpopulation model is assumed. In finite population theory, the contribution function evaluated in different subset of covariates is \textit{not} considered random. The only randomness introduced is the subsampling of coalitions. I introduce indicator variables $I_i$ where $I_i$ is the indicator function for unit $i$. Let $y_i$ denote a quantity of member $i$. The Horvitz-Thompson estimator of the population total $T$ is \citep{Horvitz01121952}
\begin{equation}
\hat{T}=\sum_{i=1}^{n_S} \frac{y_i}{\pi_i},
\end{equation}
where $\hat{T}$ is the estimator of the population total. This estimator is unbiased. I observe that each $y_i$ is weighted with the inverse of the inclusion probability. When considering the full population of units and taking the expected value of the the indicator variables, the $\pi$-terms cancel out. 

When considering regression, weighted least squares is the typical way of dealing with varying inclusion probabilities. The inverse of each of the inclusion probabilities is multiplied with the (possibly, meaning if it exists) corresponding weight in the weight matrix. If unit 1 has inclusion probability 0.1, the diagonal element of $W$ corresponding to unit 1 is divided by 0.1.

\section{Brief overview of Shapley values used to explain individual model predictions}\label{sec:BriefOverview}
This section is a slightly re-written version of Section 2 in \cite{aanes2025fastapproximativeestimationconditional}, which I wrote in April 2025. I introduce the Shapley value, how it can be used to explain model predictions and I explain how we can calculate the Shapley values. 

The Shapley value \citep{Shapley1953} originates from game theory. It is used to distribute payoff to the players of a game based on their contributions. In machine learning the typical use is to tell how much the different features contributed to a specific prediction of a model. Next, I formalize this.

I want to explain a prediction of a model $f(\bm x)$ that is trained on $\mathcal{X}=\{\bm x^{[i]},y^{[i]}\}_{i=1}^{N_{train}}$, where $\bm x_{i}$ is an $p$-dimensional feature vector, $y^{[i]}$ is a univariate response and the number of training observations is $N_{train}$. The prediction $\hat{y}=f(\bm x)$ for a feature vector $\bm x=\bm x^*$ is to be explained by using Shapley values. I let the Shapley value corresponding to feature $i$ be denoted by $\phi_i$. The Shapley value vector $\bm \phi$ is a partition of the prediction into the contribution of each of the predictors. This means $\phi_i$ tells us how much the $i$-th feature contributed to the specific prediction of a model in a positive or negative way. If it is positive, the feature contributed in a positive way while if it is negative, the feature contributed in a negative way.

To calculate $\bm\phi$ I need to define what is called the contribution function. It should resemble the value of $f(\bm x^*
)$ when only the features in coalition $S$ are known.
I use the contribution function \citep{Lundberg2017}
\begin{equation}
\label{eq:shapleyConditional}
    v(S)=E(f(\bm x)|\bm x_s=\bm x_s^*),
\end{equation}
which means I consider the expected value of $f(\bm x)$ conditioned on the features in $S$ taking on the values $\bm x
^*$. For example, when $S$ consists of the first two features, then the conditional expectation is conditional on these two features only and at the specified values. I must calculate the contributions for each possible subset of covariates to be able to find the Shapley values. For $p$ features, there are $2^p$ possible subsets, which means the procedure becomes non-doable when there are many features.

I use the so called KernelSHAP procedure so that I estimate $\bm \phi$ by solving a linear weighted least squares problem. The Shapley value vector $\bm\phi$ is obtained by solving the system of equations 
\begin{equation}
\label{eq:kernelShap}
    \bm Z^T\bm W\bm Z\bm\phi  = \bm  Z^T \bm W \bm v,
\end{equation}
here $\bm v$ is the vector of the expected contribution function of length $2^p$, $\bm Z$ is a $2^p\times (p+1)$ binary matrix (0 and 1), where the first coloumn is a coloumn vector of 1s and the $i$ element in the j-th row is 1 if covariate $i$ is  in the set $S$ corresponding to the $j$-th value of contribution function and otherwise 0, and the weight matrix $\bm W$ is a diagonal matrix  $2^p\times 2^p$ with elements $k(p,|S|)=(p-1)/\binom{N}{k}|S|(p-|S|)
$. I let $|S|$ denote the number of covariates in the set $S$. I set $k(p,0)$ and $k(p,p)$ to $10^6$, which is also done in the shapr-package \citep{jullum2025shapr}.

There are several ways to estimate $v(S)$. In my work I sample coalitions and then estimate $v$ using regression. One could instead have used Monte-Carlo sampling to approximate each conditional expectation. 
In the traditional KernelSHAP procedure, one also samples subsets (with replacement), which are also called coalitions, used to estimate $\bm \phi$. One uses the Shapley weights $k(p,S)$ as sampling weights and always include the empty and full coalitions in the estimation.  

There are several possible Shapley values. One popular Shapley value is called the marginal Shapley value. The term \textit{marginal} is due to the fact that it does not take into account feature dependence in the estimation phase. The first paper to take into account feature dependence is \cite{Aas2021}, and the corresponding Shapley value is the conditional one. The shap package \citep{Lundberg2017} is a very popular package used to estimate (mainly) marginal Shapley values, while shapr \citep{jullum2025shapr} estimates conditional Shapley values.

One can find more details on Shapley value estimation and background theory for prediction models in e.g \cite{Lundberg2017}, \cite{Aas2021} or \cite{olsen2023comparative}.

The sampling of coalitions in KernelSHAP like procedures is typically done with replacement. In some of the procedures the coalitions with highest weight are always included provided there are enough samples available. Then the remaining coalitions are sampled. In the method of \cite{olsen2025improving}, expected weights are used in the estimation. The use of expected weights instead of sampling fractions reduce the variability of the weights. I consider this method in the Shapley value estimation framework. 

In the sampling of coalitions, I obtain a sample $S$ of coalitions, so that the approximated Shapley values become
\begin{equation}
\label{eq:kernelShapApprox}
    \bm Z_S^T\bm W_S\bm Z_S\bm\phi  = \bm  Z_S^T \bm W_S\bm v,
\end{equation}
where I only include elements of $Z$ in $Z_S$ that actually correspond to sampled units. The weight matrix $W_S$ is typically modified to take into account the sampling procedure and the properties of the sampling procedure, and again only sampled units are included. 

Next I introduce the concept of pairing, see e.g Chapter 9 in \cite{kroese2013handbook}, which reduce the variance of the Shapley values. Imagine there are 6 features. When the coalition with features 1 and 2 is sampled then the coalition with features 3,4,5 and 6 is also included in the computations For applications within explainable artificial intelligence, see e.g \cite{covert2021improving},  \cite{mitchell2022sampling} or \cite{olsen2025improving}. This means we can sample half of the groups of features, and use the pairing strategy to pick the remaining coalitions. In the example we can sample coalitions for feature size 1, 2 and 3 and use the pairing strategy. Care must be taken when sampling from the 3rd group, as half of the coalitions are paired with the remaining half. We use this pairing strategy in the calculations, both for the shapr package (which is the default behavior) and for the new computational algorithm.        

To calculate the contribution function fast, I use the result in \cite{aanes2025fastapproximativeestimationconditional}. In that work a new method for fast calculations of linear models is derived. Without using multiple CPUs/GPUs, i.e by using one core, the mean over 65 000 linear models can be estimated within minutes, which is quite impressive. This means we estimate a linear model and provide predictions using the estimated model. To explain the predictions, we fit linear models to the predictions of the model. The R-script uses sparse matrices and the Cholesky decomposition in the Matrix-package \citep{MatrixBates}. Also some functions from the R-package kernelshap are re-used \citep{kernelshap}. I use R-version 4.5.2 \citep{Rcite} and also, as previously mentioned, use version 1.0.7 of shapr. 

\section{Combining finite population sampling theory with Shapley value estimation theory}\label{sec:CombineFiniteAndKernelSHAP}
I now introduce the sampling of coalitions.
The are two main steps in the sampling of coalitions. First we group the coalitions according to the number of features considered in the coalition. This means all coalitions with two features are in one group, while coalitions with 3 features are in another group and so on. Thereafter, I apply an algorithm that samples the number of coalitions in each group. After the number of coalitions in each group has been established, we sample without replacement for each group the number of coalitions that the algorithm outputted. The sampling is done so that there is no dependence between groups, except for the paired groups, which are dependent on each other. 

The algorithm that is used to calculate the number of coalitions in each group is the expected values of the Wallenius' noncentral hypergeometric distribution. I give a short overview of the theory presented in the note ``Biased Urn Theory'' which accompanies the R-package BiasedUrn \citep{Fog2024}. Here the distribution is introduced. Each group of coalitions have a weight $\omega_i$. I draw a total of $n$ coalitions across groups. The coalitions are drawn one by one in such a way that the probability that a coalition is sampled, is equal to this coalition's fraction of the total weight of all coalitions that lie in the pool of non-sampled coalitions. This distribution has the following probability mass function:
\begin{equation}
    p_W(\bm x;\bm mn, \bm  \omega)=(\prod_{i=1}^c  {m_i\choose x_i})\int_o^1 \prod_{i=1}^c (1-t^{\omega_i/d})^{x_i}\mathrm{d}t),
\end{equation}
where $d=\sum_{i=1}^c \omega_i(m_i-x_i)$, and $c$ is the number of groups of coalitions, while $\bm x=(x_1,x_2,\ldots x_c)$ is the number of coalitions drawn from each group of coalitions, $\bm m=(m_1,m_2,\ldots m_c)$ is the original/initial number of coalitions of each group of coalitions. Finally, $n=\sum_{i=1}^c x_i$ is the total number of coalitions drawn and as explained before, $c$ is the number of groups of coalitions.
There are conditions that must be fulfilled for the Wallenius' distribution to be usable. These are listed in the note and I now present them in this paper:
\begin{enumerate}
    \item Items are taken from a finite population of items that are to be sampled. The sampling is without replacement.
    \item Items are drawn one by one. However, in the function one can specify several draws.
\item The total number of balls to be drawn, $n$, is fixed and is not dependent on which items that are already taken.
\item The probability of an item to be sampled at a particular draw,  is equal to this item's fraction of the total weight of all items that have not been sampled at that moment. The weight does only depend on which group/kind it belongs to.
\end{enumerate}

Instead of sampling the number of coalitions of each group, I calculate expected values and round these to the nearest number. This will reduce variability in the weights.  

If the expected number of samples units from the first group is e.g 16 and there are 16 coalitions, then the inclusion probability is 1, as explained in Section \ref{sec:FinitePop}. If there are 100 coalitions in group 2 and the expected number sampled is 300, then the first order inclusion probability of each of the 100 sampled coalitions is 100/300, which is 1/3.
\subsection{Comparison with KernelSHAP}
In KernelSHAP the sampling is done with replacement. Due to the sampling design, where coalitions with lowest weight are sampled less often than coalitions with higher weight, I obtain a weighted version of the Coupon Collector's Problem. The problem first appeared in \textit{De Mensura Sortis} (On the Measurement of Chance) written by A. De Moivre in 1708. In the weighted version, different probabilities are assigned to collect each unique coupon. This makes rare coupons harder to find than more common ones. The similarity between KernelSHAP and the Coupon Collector's  Problem (CC-problem) is due to the way the items are drawn and the varying probabilities of the diffent outcomes. Once you have collected a copuon, you can in another trial draw the same coupon. This means one can think that the items are collected with replacement, and since we have different probabilities, the two problems are equivalent. In KernelSHAP we want to draw more and more unique coalitions, and this becomes harder and harder due to the probability weighting. With the new approach, the CC-problem can be avoided, as you simply specify the number of unique balls to draw, and the algorithm calculates the distribution of the balls into the different categories/strata. In the new procedure, the \textit{remaining} balls compete to be drawn. I calculate the average number of balls drawn in each category/strata, so that the algorithm is completely deterministic when it comes to the distribution of balls/coalitions, but random when it comes to which coalitions are sampled within each category/strata (provided I do not sample every coalition in a category).

\section{Variance estimation of Shapley values}\label{sec:VarEstimation}
I introduce three Shapley value variance estimators, and then argue why the first should not be used. The main reason is incorrectness.

\subsection{Necessary conditions for validity of bootstrap methods for finite populations}
One way of obtaining conditions for a bootstrap method so that it is valid, is to equate the moments (mean value, variance and covariance)  with the Horvitz-Thompson estimator. More details follow next. Let $S_k$ denote the number of times unit $k$ is included in the sample. The HT estimator of the total for a single bootstrap sample is, see e.g \cite{antal2014new}
\begin{equation*}
    \hat{Y}=\sum_{k\in S}\frac{y_k}{\pi_k}S_k.
\end{equation*}
It then follows that 
\begin{equation*}
    \mathrm{E}(\hat{Y})=\sum_{k\in S}\frac{y_k}{\pi_k}E(\mathrm{S_k}),
\end{equation*}
and 
\begin{equation*}
    \mathrm{var}(\hat{Y})=\sum_{k\in S}\sum_{l\in S}\frac{y_k}{\pi_k}\frac{y_l}{\pi_l}\textrm{cov}(S_K,S_l). 
\end{equation*}
If we want the expected value of $\mathrm{E}(\hat{Y})$ to equal the HT estimator of the total, then we must have 
\begin{equation}
    E(S_k)=1, k\in S.
\end{equation}
Furthermore, if we want the bootstrap estimator to provide an unbiased estimator of the HT total estimator, two conditions are necessary and sufficient, namely
\begin{equation}
    \mathrm{var}(S_k)=1-\pi_k, k\in S,
\end{equation}
and 
\begin{equation}
    \mathrm{cov}(S_k,S_l)= 1-\frac{\pi_k\pi_l}{\pi_{k.l}}.
\end{equation}
When considering the weighted least squares problem in KernelSHAP, we have two HT-like estimators, where we divide $Z^TWv=\sum_{i\in S}Z^Tw_iv_i$, the first HT like estimator, by $Z^TWZ=\sum_{i\in S}Z_i^Tw_iZ_i$, the second HT like estimator. The matrix $Z^TWZ$ typically varies less than the other HT like estimator. This means that it is natural to impose the above restrictions on the new bootstrap method as well.

\subsection{Horvitz-Thompson like estimator}

I study next a variance estimator, which I later explain is incorrect. It is intuitive, but should not be used. 

In the expression for the Shapley values, there are two Horvitz-Thompson like estimators. The first is $P_1=Z_S^T W_SZ_S$, which can be rewritten as $\sum_i Z_i^T w_{ii}Z_i$ and the other one is $Z^TW\bm v$, which can be rewritten as $\sum_i Z_i^T w_{ii}v_i$. It can be shown that $E(P_2)=Z^TWZ$. I can in fact calculate $Z^TWZ$ without calculating the contribution function for each of the coalitions. I assume I have sampled enough coalitions so that $P_1$ has very low variance. Then I only need to take into account the variance of $P_2$ when calculating the variance of the Shapley values, as I have assumed $P_1$ is (close to) deterministic.

The variance of the Horvitz-Thompson estimator for the total is:
\begin{equation}
   \textrm{Var}(\hat{T})=\sum_{k\in U}\sum_{l\in U}\textrm{Cov}(I_K,I_l)\frac{y_k}{\pi_k}\frac{y_l}{\pi_l}
\end{equation}
and for a sample $S$ an unbiased estimator for $\textrm{Var}(\hat{T})$ is 

\begin{equation}
   \widehat{\textrm{Var}}(\hat{T})= \sum_{k\in S}\sum_{l\in S}\frac{\textrm{Cov}(I_k,I_l)}{\pi_{kl}}\frac{y_k}{\pi_k}\frac{y_l}{\pi_l}=\sum_{k\in U}\sum_{l\in U}I_kI_l\frac{\textrm{Cov}(I_k,I_l)}{\pi_{kl}}\frac{y_k}{\pi_k}\frac{y_l}{\pi_l}
\end{equation}
We have unbiasedness as $\textrm{E}(I_kI_l)=\pi_{kl}$ and the only stochastic part is $I_kI_l$.

In our design, the strata are paired, meaning that two and two strata are dependent and each pair is independent from the remaining pairs of strata. This simplifies calculations, but the procedure is still computationally expensive due to the double summation. When performing calculations, I use that $\pi_{kk}=\pi_k=n/N$ and that $\textrm{Cov}(I_k,I_l)=\pi_{kl}-\pi_{k}\pi_{l}$. The indicator function for each coalition in a pair of coalitions are equal to each other. This enables us to calculate the covariances.

The above method is not correct. If we use the total value $\sum_{i\in U}Z^T_iw_iZ$, where $U$ denotes the universe with all possible coalitions, there is no difference between using paring and not by using pairing when considering this matrix. Explained differently, $\sum_{i\in S}Z_i^Tw_iZ_i$ takes different values for pairing and when not by using pairing. By not using pairing, I mean we sample each group of feature size independently of the other groups. Therefore I do not consider this method in the numerical studies.

\subsection{The Doubled half bootstrap}

In \cite{antal2014new} the authors propose several bootstrap methods. We focus on a very fast method of bootstrapping for sampling without replacement
from a finite population. It does not require artificial populations or bootstrap weights. The bootstrap samples are chosen directly 
from the original sample. The procedure contains two steps: in the first step,
units are selected once using Bernoulli sampling with the same inclusion probabilities
as in the original design. In the second step, amongst the non-selected units, half of the
units is randomly selected twice. There are some details, which are specific to simple random sampling without replacement, and these can be found in section 6 in the paper. The theory is presented for simple random sampling, but in our stratified design, the strata are paired. This means that I can apply the algorithm for one of the strata in  each pair of strata and use the pairing requirement to find the corresponding coalition when sampling coalitions.

\subsection{The Symmetric bootstrap method}
In the new method, each member of the original sample can be present 0, 1 or 2 times in each bootstrap sample. I name it the Symmetric bootstrap as the number of times 0 or 2 occurs in total in each sample are equal. In brief, the new method is as follows, assuming integer values in step 1:
\begin{enumerate}
    \item Pick the number of times 0, 1 and 2 occurs in a specific stratum.
    \item Repeat several times for each stratum.\begin{enumerate}
        \item
    Draw without replacement from the original sample, where one chooses which element belongs to each of the categories 0, 1 or 2. 
    \item Calculate the Shapley values after performing the previous step for each stratum.  
\end{enumerate}
\end{enumerate}

To find the number of times 0, 1 or 2 occurs in total in each bootstrap sample, one uses formulas. One calculates the expected value using the probabilities in Equation \eqref{eq:pr0and2} and Equation \eqref{eq:prob1}, to be introduced later. If $n=20$ and $N=40$, then the expected number of 0's and 2's is $0.5\cdot 20\cdot(1-20/40)$, which is 5, and the expected number of 1's is $20\cdot20/40$ which is 10. In this case the algorithm is exact. However, if I obtain fractional values, the algorithm is approximate. I then need to introduce a Bernoulli random variable that alternates (possibly unevenly) between choosing two consecutive values of $n_2$. Details are given in Section \ref{sec:NonInteger}.

\subsubsection{Verifying the correctness of the new method}
In the new bootstrap procedure, each random variable can take three values, 0, 1 or 2 with probabilities
\begin{align}
    \textrm{Pr}(S_k=0)=& \textrm{Pr}(S_k=2)=\frac{1}{2}(1-\frac{n}{N})\label{eq:pr0and2}\\
\textrm{Pr}(S_k=1)&=\frac{n}{N}\label{eq:prob1}
\end{align}
It is  easily verifiable that $\sum \textrm{Pr}(S_k=i)=1$ and that the expected value of $S_k$ is 1. The variance of $S_k$ is 
\begin{align}
    \textrm{Var}(S_k)&=\textrm{E}(S_k-\mu)^2=(0-1)^2\cdot \textrm{Pr}(S_k=0)+(1-1)^2\cdot \textrm{Pr}(S_k=1)+(2-1)^2\cdot \textrm{Pr}(S_k=2)\\
    &=2\cdot \textrm{Pr} (S_k=2) = 1-\frac{n}{N},
\end{align}
which is the correct expression. 

I also derive an alternative expression for the variance. I let $N_2=2n_2+n_1$ denote the total number of items, where $n_2$ is the number of 2's and since the number of 0 and 2's are equal, I obtain $2n_2$. I have that $N_2=n$, where $n$ is the number of items drawn originally. 
I find that $\textrm{Pr}(S_k=2)=\frac{n_2}{n}$ as the number of 2's is $n_2$ and I have $n$ items to choose from. I therefore obtain the equivalent expression $\textrm{Var}(S_k)=\frac{2n_2}{n}$.

Now I would like to derive an expression for the covariance between $S_k$ and $S_k$ for $k\neq l$. I then need the joint inclusion probabilities. I draw the items in a special way. I draw the category 2 a number of $n_2$ times and category 1 a number of $n_1$ times. I therefore do \textit{not} perform independent draws. I have that $\textrm{Pr}(S_k=1,S_l=1)=\frac{n_1(n_1-1)}{n(n-1)}$, that $\textrm{Pr}(S_k=2,S_l=2)=\frac{n_2(n_2-1)}{n}$ and that $\textrm{Pr}(S_k=2,S_l=1)=\textrm{Pr}(S_k=1,S_l=2)=\frac{n_1n_2}{n(n-1)}$. By definition $\textrm{Cov}(S_k,S_l)=\textrm{E}((S_k-1)(S_l-1))=\textrm{E}(S_kS_l)-1.$ By definition I have that 
\begin{align}
\textrm{E}(S_kS_l) =& 0\cdot \textrm{Pr}(S_k=0,S_l=0)+0 \cdot \textrm{Pr}(S_k=1,S_l=0)+0\cdot \textrm{Pr}(S_k=1,S_l=0)+\\ &\textrm{Pr}(S_k=1,S_l=1)+2 \textrm{Pr}(S_k=1,S_l=2)+2 \textrm{Pr}(S_k=2,S_l=1)+4 \textrm{Pr}(S_k=2,S_l=2).
\end{align}
By inserting the joint inclusion probabilities I obtain that \begin{equation}
  \textrm{Cov}(S_k,S_l)  =-2n_2/n(n-1)=-\textrm{Var}(S_k)/(n-1), 
\end{equation}which is correct. I note that I have assumed integer values of $n_1$, $n_2$ and $n_0$. If this is not so, see Section \ref{sec:NonInteger}.

\subsubsection{Non-integer $n_0$, $n_1$ or $n_2$}\label{sec:NonInteger} If $n_0$, $n_1$ or $n_2$ are non-integer, I have to alter the procedure and introduce a Bernoulli random variable, as explained next. The formulas for the variance and covariance have one free parameter, namely $n_2$. I  see that both formulas depend on $n_2$ linearly, while $N$ and $n$ are fixed. If the variance formula is satisfied, then automatically the covariance formula is satisfied and vice versa. If I solve for $n_2$ in the variance formula, I obtain 
\begin{equation}
    n_2=\frac{1}{2}(n-n^2/N)=\frac{1}{2}n(1-n/N),
\end{equation}
which I identify as the contribution of $n_2$ to the expected value. If this expression is correct on average, I obtain approximately the correct expected value of 1 of $S_k$. In addition I obtain approximately the correct variance and covariance formulas. Consider an example application where $n=5$ and $N=13$. Now I calculate $0.5\cdot 5\cdot (1-5/13)$ which equals 1.538. So, if one in e.g 500 iterations uses $n_2=2$ in $53.8\%$ of the iterations, and the remaining $46.2\%$ of the iterations one uses $n_2=1$, then $n_2$ is approximately equal to the correct expected value of $n_2$. I introduce a Bernoulli random variable that chooses which value of $n_2$ to use, with success probability as described. This means I can obtain a fully automatic boostrap procedure for all choices of $n$ and $N$.  

\subsubsection{Comparison with other methods}
In \cite{Zeinab} an overview of bootstrap methods for finite populations is given. They also explain how to obtain bootstrap estimates of the standard deviation. The earliest work on pseudo-population methods is \cite{gross1980median}. In that work one assumes that $N/n$ is integer. Next a pseudo population, $U^*$, of size $N$ is created by duplicating $N/n$ times each unit $k$ in the original sample. Then a sample of size $n$ is chosen from $U^*$ without replacement. When $n$ is non-integer, there are several proposed solutions, such as \cite{Booth01121994}. In the new, current work I mimic the sampling procedure by sampling without replacement from the original sample. The method of \cite{antal2014new} is also very shortly discussed. I  have chosen to include it because it is fast and there exists an implementation in the R-package svrep \citep{Schneider2025}, where the function   \textit{make\_doubled\_half\_bootstrap\_weights} is used. One has to e.g $output = "ww"$ to output the number of replications of each member in the original sample. 

\subsection{Variance estimation in shapr}
In shapr there is a built-in bootstrap method that is used when the iterative procedure is used. With the iterative method, the coalitions are sampled in batches until a convergence criterion is fulfilled. In one run of the algorithm, for instance first 5 coalitions are sampled, then 15 and finally 14 coalitions are sampled because the criterion finally is fulfilled. The details about the bootstrapping method implemented in shapr is scarce in the reference paper \cite{jullum2025shapr}, but it contains a reference to Section 4.2 in \cite{goldwasser}. Here the authors propose to take $M$ samples with replacement from the observed subset $K$ of coalitions and thereafter calculate the KernelSHAP weights. The procedure is redone several times to obtain several bootstrap estimates.  

I force the iterative method to never finish before using all coalitions by setting the value of the convergence tolerance to 0. I however limit the number of coalitions used by the method by setting a maximum value of 16 coalitions in the small study and 402 in the larger simulation study, including the empty and grand coalitions, so that the algorithm in total will use 16 (or 402) coalitions in every run and never finish before. One example call to shapr is given below.   
\begin{lstlisting}[language=R]
xplain_separate_all_iterative <- shapr::explain(
  model = fit,
  x_explain = x_train,
  x_train = x_train,
  approach = "regression_separate",
  phi0= p1,
  regression.model = parsnip::linear_reg(),
  iterative = TRUE,
  seed=i+1,max_n_coalitions = 16,iterative_args = list(convergence_tol = 0),
)
\end{lstlisting}
In the above code I vary the seed from run to run as I iterate over the variable \textit{i}.

\section{Simulation studies}\label{sec:SimStudies}
\subsection{Smaller simulation study}\label{sec:Smallstudy}

In the smaller simulation study I pick the variables Infant deaths, Under five deaths, GDP per capita, Thinness five nine years and Schooling as features while the response variable is Life expectancy. I have here presented the short names of the variables. The first half of the observations, namely observation 1 to observation 1432, are used to train the linear model, while I explain the remaining half using Shapley values. 

Care is needed when comparing methods against each other. For instance, one could pick one sample and then apply bootstrapping using this sample using e.g. the Symmetric bootstrap method. Thereafter one could compare against the bootstrap estimates from one of the runs of the shapr package. However, the result could be terribly misleading due to the specific sample of coalitions one applies bootstrapping to. Instead I suggest to do the following: Run each method 300 times. Then calculate the corresponding 300 Shapley value means. One can use the standard deviation function on these means. These values will correspond to resampled estimates since one samples from all available coalitions. In my small study I sample 14 coalitions in total, and in addition include the grand and empty coalitions. To find bootstrap estimates for each of the 300 samples, I perform bootstrapping using the corresponding sample as sampling space in the bootstrap procedure. Then, one takes the averages of the bootstrap estimates, say 300 or 400. One obtains 300 such means of standard deviations, as there are initially 300 samples of coalitions. I bootstrap also 300 times, but I could have chosen another number.

\begin{figure}
\centering
\includegraphics[width=0.75\textwidth,trim=0cm 0cm 0 0cm, clip,center]{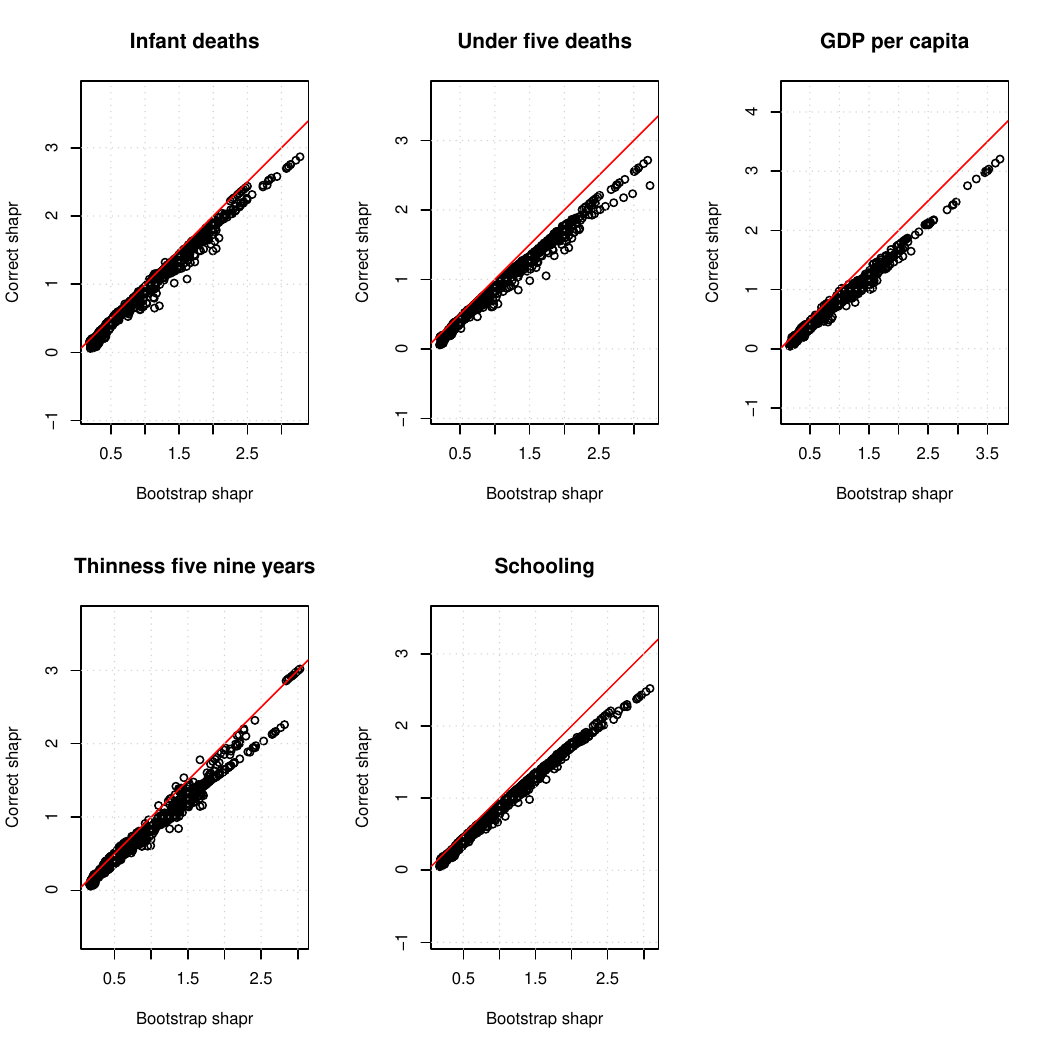}
\caption{
Small study: The bootstrap estimates of the standard deviations obtained using shapr are plotted against the standard deviation based on running shapr 300 times, finding the mean and taking the standard deviation. Both estimates are based on the same runs of shapr. There are more details provided in the text. Along the x-axis is the bootstrap estimates while along the y-axis is the true standard deviations. }\label{fig:Small_StudyNR}
\end{figure}

In Figure \ref{fig:Small_StudyNR} I plot the results using shapr and compare bootstrap estimates against estimates from resampling. I call the resampled estimates for true values as correct sampling is applied, namely using 16 coalitions and sampling from all possible coalitions. We observe that shapr can be a little more conservative than the true values, meaning sometimes the bootstrap procedure outputs too large standard deviations.         

\begin{figure}
\centering
\includegraphics[width=0.75\textwidth,trim=0cm 0cm 0 0cm, clip,center]{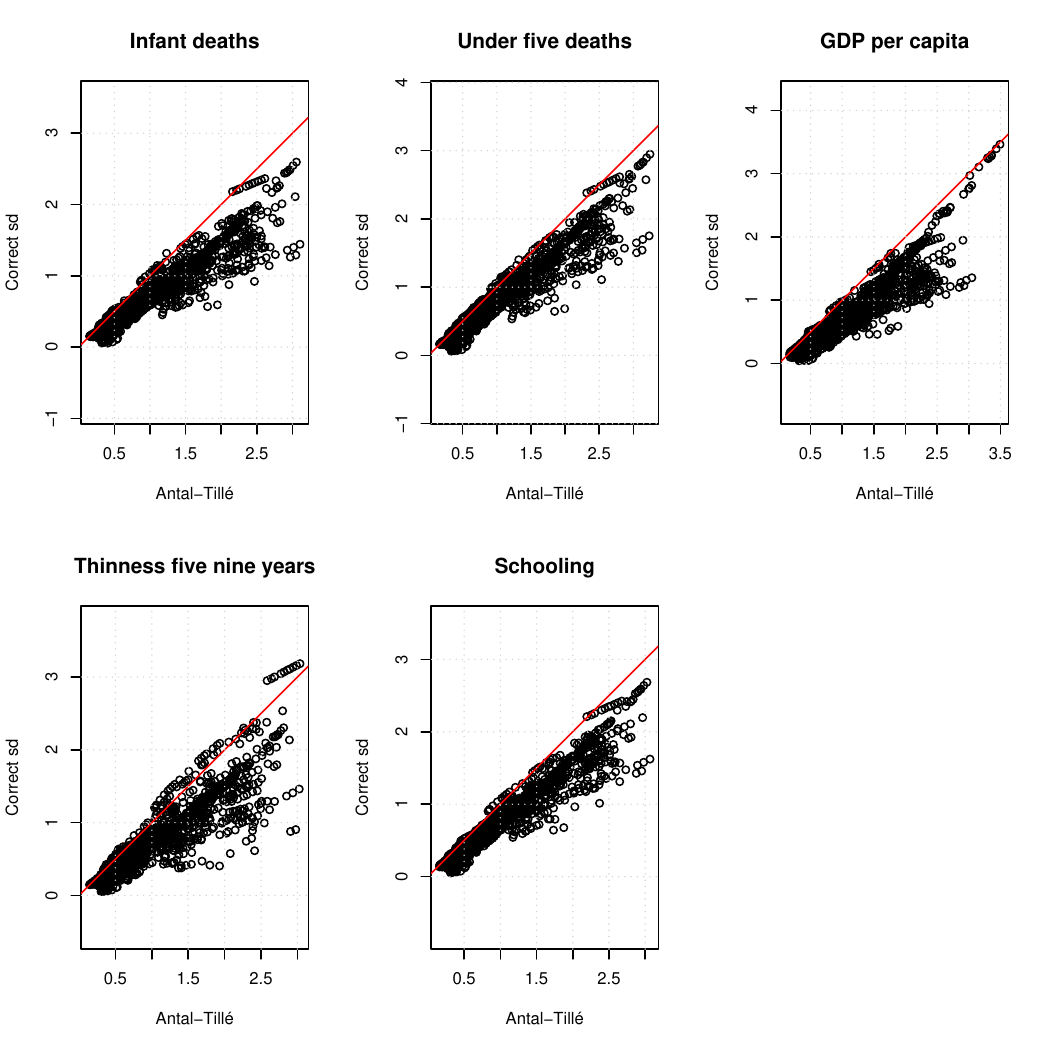}
\caption{Small study: The bootstrap estimates of the standard deviations obtained using the new method for the sampling of coalitions combined with the Antal-Tillé bootstrap method are plotted against the standard deviation based on running the new algorithm 300 times, finding the mean and taking the standard deviation. Both estimates are based on the same runs of the new algorithm. More details are given in the text. Along the x-axis is the bootstrap estimates while along the y-axis is the true standard deviations. }\label{fig:SmallStudyAntalTille_vs_true}
\end{figure}

\begin{figure}
\centering
\includegraphics[width=0.75\textwidth,trim=0cm 0cm 0 0cm, clip,center]{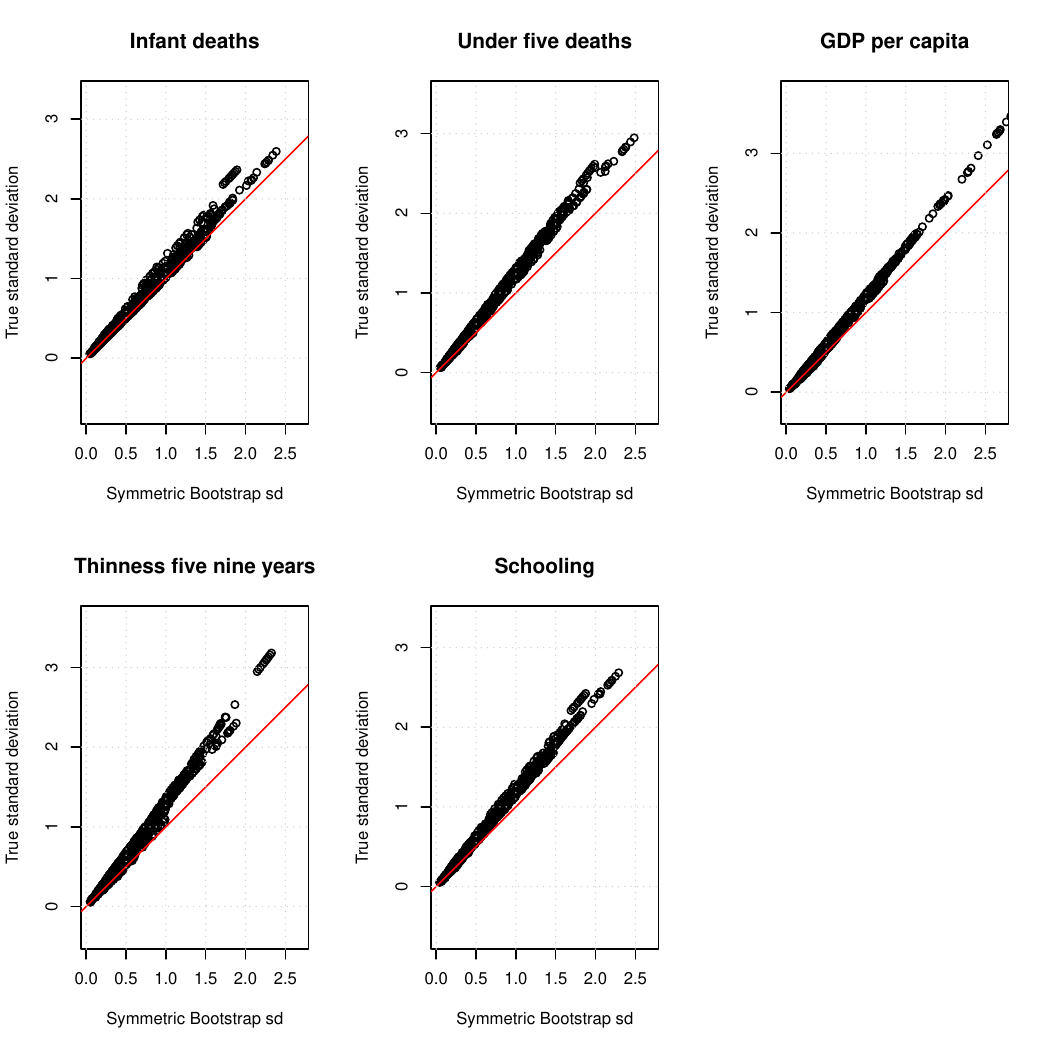}
\caption{
Small study: The bootstrap estimates of the standard deviations based on the Symmetric bootstrap obtained using the new method for sampling coalitions are plotted against the standard deviation based on running the new algorithm 300 times, finding the mean in each run and taking the standard deviation. Both estimates are based on the same runs of the new algorithm. More details are provided in the text. Along the x-axis is the bootstrap estimates while along the y-axis is the true standard deviations. }\label{fig:SmallstudyNewMethodFredrikvs_true}
\end{figure}

In Figure \ref{fig:SmallStudyAntalTille_vs_true} the bootstrap estimates from the Antal-Tillé-method are quite poor when one compares the plot to the plot with shapr. One problem with the bootstrap procedure is that the rank of $\bm Z_S^T\bm W_S\bm Z_S$ is not always full-rank. We have then removed this bootstrap estimate of $NA$-values from the calculations afterwards. Out of 300 bootstrap iterations, on average (i.e of the initial 300 samples) in 50 iterations we obtain a matrix which is not full-rank, which is quite high. The problem of $ZWZ$ not being full rank is also an issue for the Symmetric bootstrap. However, now only 15.81 on average bootstrap sample give a $ZWZ$-matrix which is not full rank. We see from Figure \ref{fig:SmallstudyNewMethodFredrikvs_true} that the quality of the bootstrap estimates are much better than for the Antal-Tillé-method instead. The bootstrap algorithm can be a little liberal, i.e produce somewhat smaller standard deviations than what is obtained through resampling.  

\begin{figure}
\centering
\includegraphics[width=0.75\textwidth,trim=0cm 0cm 0 0cm, clip,center]{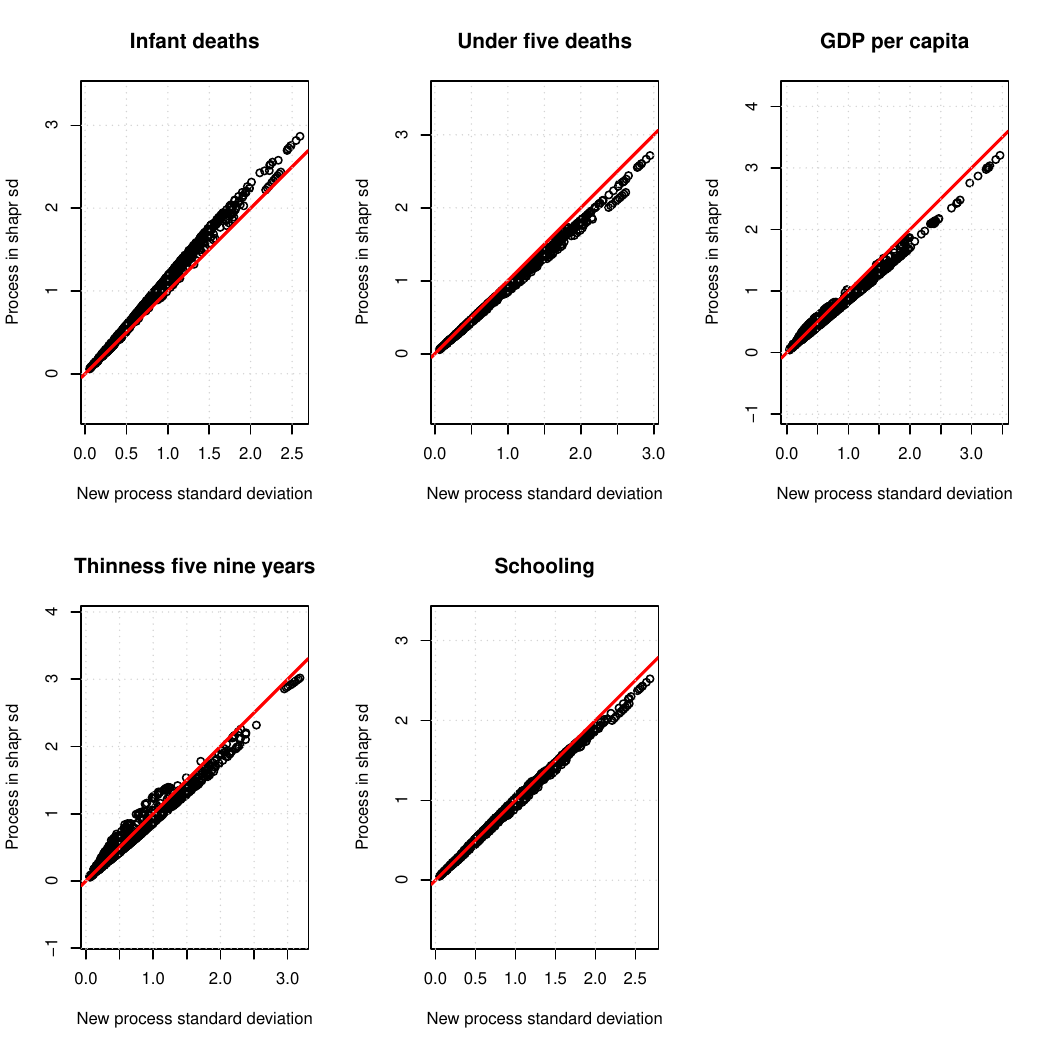}
\caption{
Small study: The estimates of the standard deviations based on resampling 300 coalitions using either the new or the shapr-method are plotted against each other. Along the x-axis is the estimates using the new method, while along the y-axis is the estimates from using the shapr-package. }\label{fig:SmallStudyNewVSshaprSD}
\end{figure}

In Figure \ref{fig:SmallStudyNewVSshaprSD} I compare the resampling estimates using the new sampling procedure against shapr.  For instance for Infant deaths, the shapr package produces a little bit larger standard deviations than the new method, while the opposite is true for the GDP per capita variable. On average the two methods, namely shapr and the new method of sampling coalitions without replacement, perform similarly. However, the bootstrap procedures differ, where the one in shapr is a little conservative while the Symmetric bootstrap can be a little liberal.    

\begin{figure}
\centering
\includegraphics[page=1,width=0.65\textwidth,trim=0cm 0cm 0 0.cm, clip,center]{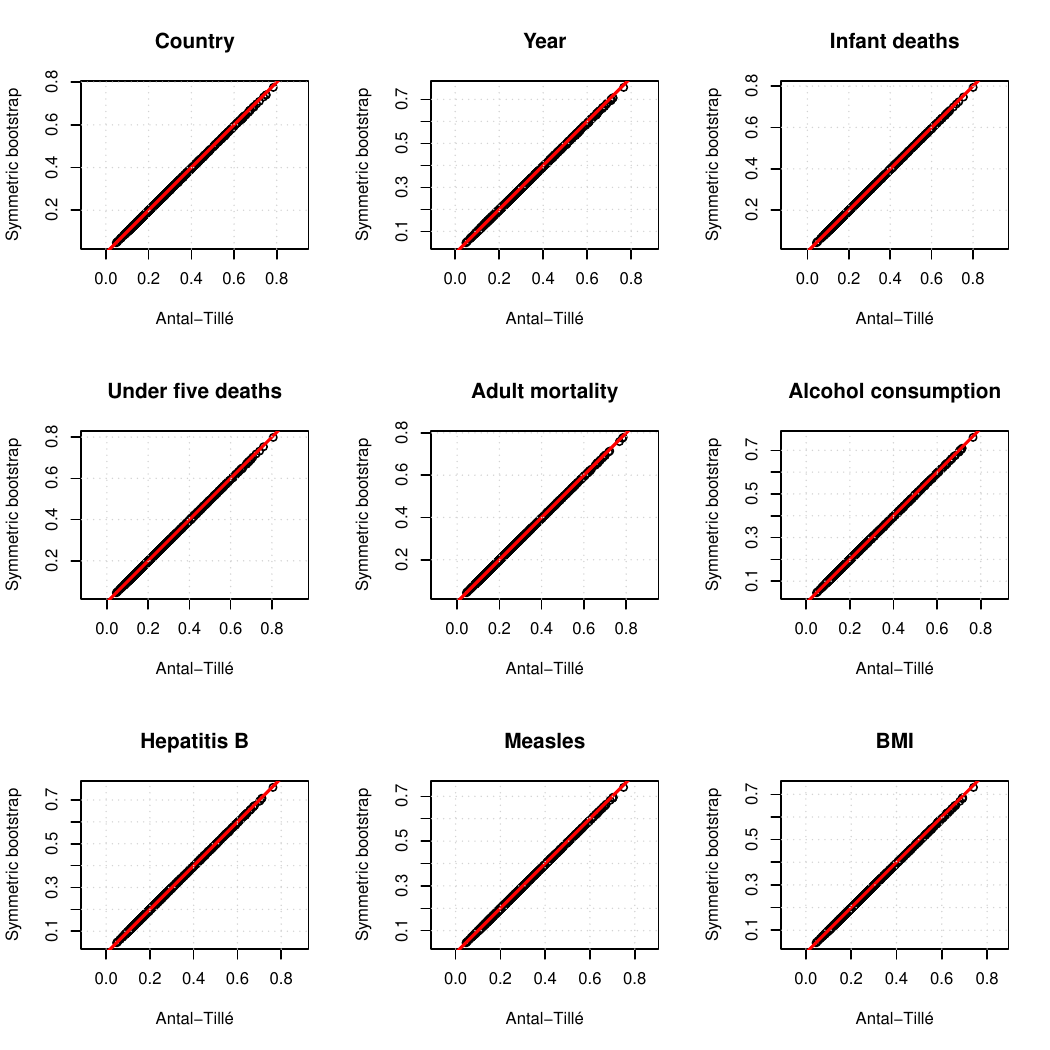}
\includegraphics[page=2,width=0.65\textwidth,trim=0cm 0cm 0 0cm, clip,center]{AntalTilleFredrik.pdf}
\caption{Large study:
Bootstrap estimates of the standard deviations obtained either through Antal-Tillé (x-axis) or the Symmetric bootstrap (y-axis.)}\label{fig:LargeStudySymmetricVsAntalTelle}
\end{figure}

\begin{figure}
\centering
\includegraphics[page=1,width=0.65\textwidth,trim=0cm 0cm 0 0.cm, clip,center]{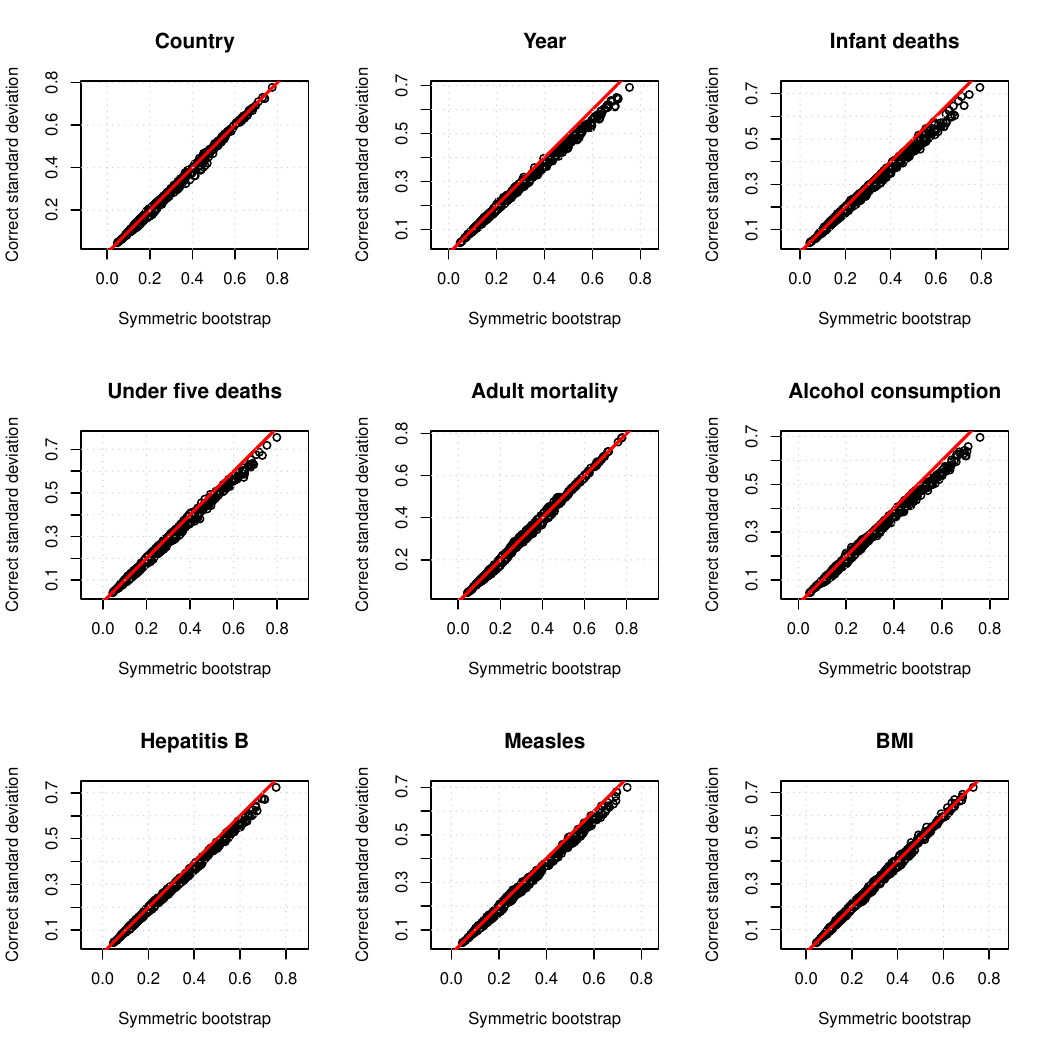}
\includegraphics[page=2,width=0.65\textwidth,trim=0cm 0cm 0 0.cm, clip,center]{Fredrik_vz_True.pdf}
\caption{Large study: The bootstrap estimates of the standard deviations obtained by using the Symmetric bootstrap method are plotted against estimates from full resampling using all available coalitions in the sampling step.}
\label{fig:LargeStudySymmetricBootVsTrue}
\end{figure}

\begin{figure}
\centering
\includegraphics[page=1,width=0.65\textwidth,trim=0cm 0cm 0 0.cm, clip,center]{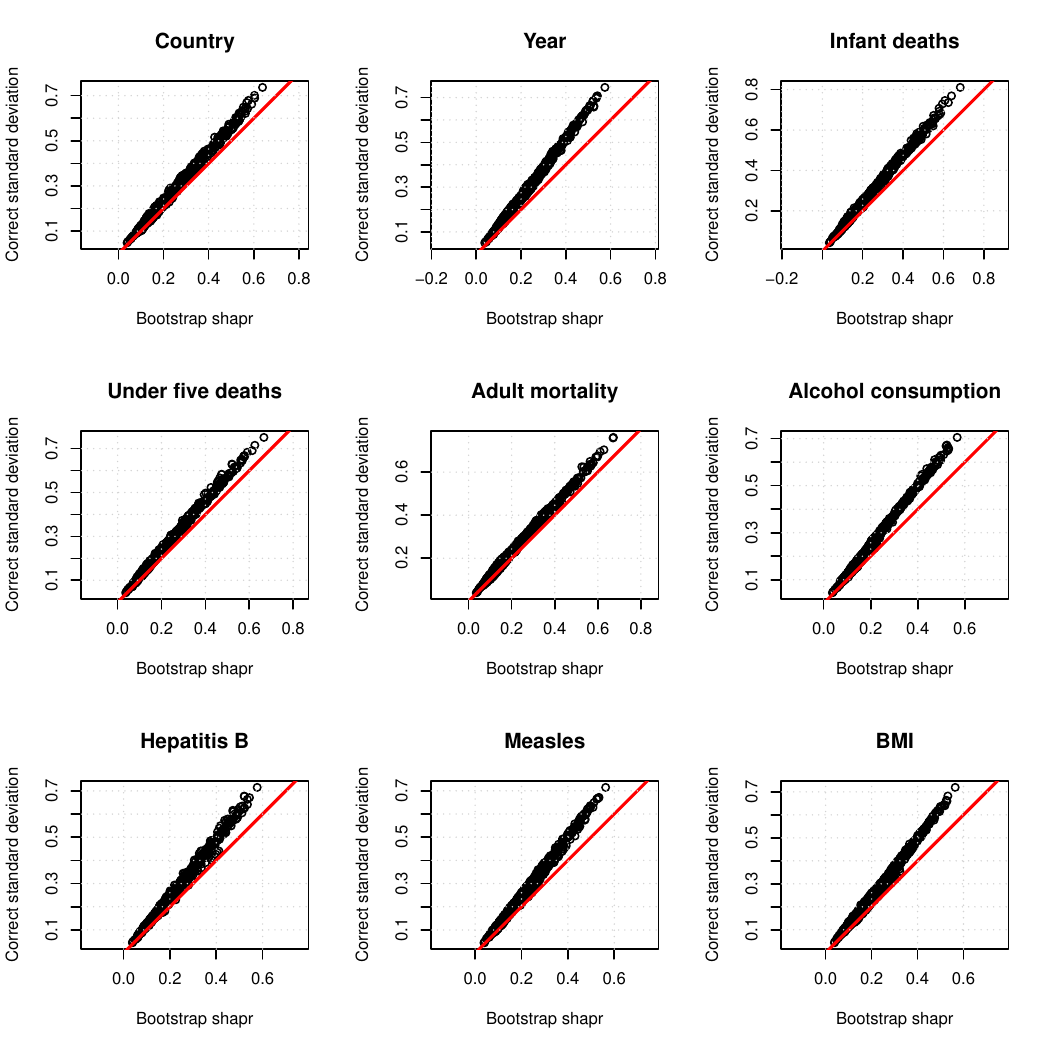}
\includegraphics[page=2,width=0.65\textwidth,trim=0cm 0cm 0 0.cm, clip,center]{NRlarger.pdf}
\caption{Large study: I plot the mean of 300 bootstrap estimates of the standard deviation against the standard deviation of 300 resamples using all coalitions. In this plot the bootstrap estimates from shapr are along the x-axis, while along the y-axis we find the estimates from taking the standard deviation of the 300 means from the runs of shapr.}
\label{fig:LargeStudytrueVsBootshapr}
\end{figure}

\begin{figure}
\centering
\includegraphics[page=1,width=0.65\textwidth,trim=0cm 0cm 0 0.cm, clip,center]{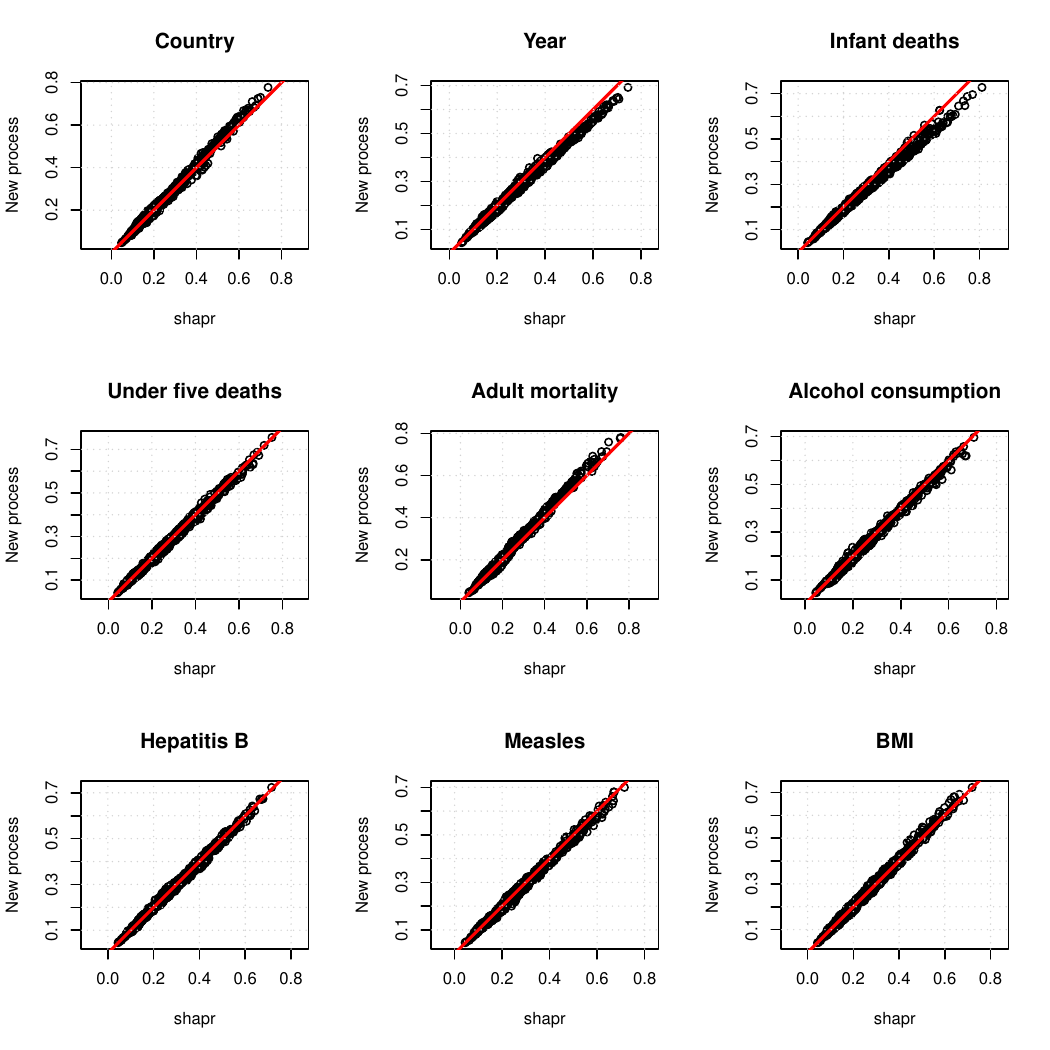}
\includegraphics[page=2,width=0.65\textwidth,trim=0cm 0cm 0 0.cm, clip,center]{NRnew.pdf}
\caption{Large study: I plot along the y-axis the standard deviations from resampling the new method while along the x-axis is the estimates from resampling by using the shapr-package.}
\label{fig:LargeStudyNewVsTrueShapr}
\end{figure}

\subsection{Larger simulation study}
In this section I perform a similar study as in Section \ref{sec:Smallstudy}. However, now I include 16 variables, namely Country, Year, Infant deaths, Under five deaths, Adult mortality, Alcohol consumption, Hepatitis B, Measles, BMI, Polio, Diphtheria, Incidents HIV, GPD per capita, Thinness ten nineteen years, Thinness five nine years and Schooling. The provided names are the short names for the variables. I perform similar calculations as before. Now I sample 400 coalitions, and include in addition the full and empty coalitions. The bootstrap estimates of the standard deviation for the new Symmetric bootstrap method and for the Antal-Tillé method are plotted against each other in Figure \ref{fig:LargeStudySymmetricVsAntalTelle}. The estimates are indistinguishable. Since these two methods provide very similar estimates, I compare the result from resampling against the Symmetric bootstrap estimates, but could have plotted the Antal-Tillé method. The plots are given in Figure \ref{fig:LargeStudySymmetricBootVsTrue}. The estimates are very similar. In Figure \ref{fig:LargeStudytrueVsBootshapr} I plot the bootstrap estimates from the bootstrap built-in procedure in shapr against the resampled estimates. The estimates from the built-in procedure underestimates the standard deviations for most of the variables, where the underestimation is not that pronounced for the Adult mortality variable or Incidents HIV, but more pronounced for the Year variable. In Figure \ref{fig:LargeStudyNewVsTrueShapr} 
I compare the estimates obtained by resampling shapr package against the estimates from resampling the new method. We observe very good correspondence, meaning that the two methods in this case are equally good. The bootstrap methods of Antal-Tillé and the symmetric bootstrap perform a little bit better than the bootstrap method in shapr.

\section{Brief, general discussion and conclusion}\label{sec:conclusion}

The new suggested algorithm to group the features, the Wallenius' noncentral hypergeometric distribution, fits well with the KernelSHAP framework, as it samples without replacement and uses the weights to determine the number of coalitions for each group. In each draw of the process, the coalitions compete to be next coalition that is drawn. Only the remaining coalitions can be drawn. By using the expected values, we reduce the variability in the weights. 

I have also applied the standard correction of varying inclusion probabilities in survey theory to the weight matrix in KernelSHAP (which means I divided the weights in the weight matrix by the inclusion probabilities) and, combined with Wallenius' distribution, with great success.

In the small simulation study the method of Antal-Tillé performed quite poorly. This is most likely due to taking the double of half of the original sample, and that my method gives bootstrap samples with more variability. The pairing also comes into play as the two paired coalitions are linearly related. The sum of two such rows is (2, 1, 1, 1) if we consider 3 variables. We also always include (1,1,1,1). The new bootstrap method is easy to implement, easy to understand and computationally fast. To my best knowledge, there is no go-to bootstrap method when the original sample has been obtained by using random sampling without replacement. There are several procedures suggested in the literature, but several of them have not been implemented for variance estimation and can be quite complex, introducing e.g artificial populations. I strongly believe that my method can fulfill this task. My method, the Symmetric bootstrap, has been able to estimate the variance, however it produced a little too small standard deviations in the smallest simulation study for some of the variables, but always for the largest standard deviations for a specific variable. For the largest simulation study, the new bootstrap algorithm performed well.   

When we compare the plots of standard deviations obtained through resampling, where all coalitions are available, the new computational method show similar performance as the method in shapr in both simulation studies. We therefore observe that the new method by itself is equally good as the method in shapr. We have now not considered the plots of the bootstrap estimates. The bootstrap methods we have considered can be replaced by other bootstrap methods, and it is therefore I have divided the analysis into two parts, where part (1) is pure resampling and part (2) is the comparison of bootstrap estimation methods. The bootstrap methods are used instead of resampling in real world applications as it is too computationally costly to calculate the union of all the contribution functions in each resample. In my case studies I used the method of \cite{aanes2025fastapproximativeestimationconditional} to quickly calculate the contribution functions. And in real life the method is to be used when the model has been estimated by another technique than the linear model estimation method or a generalized linear model (as the predictions are linear on the transformed scale, e.g linear on the logit-scale), since my method covers these models. 

I have used the value of three hundred bootstrap resamples, which should be enough to estimate the standard deviation. If I wanted e.g confidence intervals, more samples should have been considered. 

I have chosen to consider two different setups where two different numbers of covariates are considered, one small study and another much larger study. In real life, one applies the sampling algorithm only to the larger problem as we can calculate exactly the Shapley values in the smaller case study fast. However, when we evaluate algorithms, we should consider different scenarios and see how the algorithms perform in these scenarios. 

All in all, based on the case studies, the new algorithm performs equally well as the estimation method in shapr. The new bootstrap algorithm, the Symmetric bootstrap, is very computationally fast, easy to understand and performs well.  

\bibliographystyle{chicago}
\bibliography{refererences}

\end{document}